\documentclass[a4paper,12pt]{article}
\usepackage[utf8x]{inputenc}
\usepackage{amsmath,amssymb}
\usepackage[dvips]{graphicx}

\usepackage{epsfig,authblk}
\usepackage{bm,fullpage}

\title{Patchiness and Demographic Noise in Three Ecological Examples} \author[1]{Juan A. Bonachela\footnote{Corresponding author: jabo@princeton.edu}}\author[2]{Miguel A. Mu\~noz}\author[1]{Simon A. Levin} \affil[1]{Department of Ecology and Evolutionary Biology,   Princeton University, Princeton, NJ, 08544-1003, USA} \affil[2]{Instituto Carlos I de F{\'\i}sica Te{\'o}rica y Computacional, Facultad de Ciencias, Universidad de Granada, 18071 Granada, Spain}

\date{}

\begin{document}

\maketitle

\def\longrightharpoonup{\relbar\joinrel\rightharpoonup}
\def\longleftharpoondown{\leftharpoondown\joinrel\relbar}

\def\longrightleftharpoons{
  \mathop{
    \vcenter{
      \hbox{
	\ooalign{
	  \raise1pt\hbox{$\longrightharpoonup\joinrel$}\crcr
	  \lower1pt\hbox{$\longleftharpoondown\joinrel$}
	}
      }
    }
  }
}

\newcommand{\rates}[2]{\displaystyle\mathrel{\longrightleftharpoons^{#1\mathstrut}_{#2}}}

\def\longrightleftarrows{
  \mathop{
    \vcenter{
      \hbox{
	\ooalign{
	  \raise2pt\hbox{$\longrightarrow\joinrel$}\crcr
	  \lower2pt\hbox{$\longleftarrow\joinrel$}
	}
      }
    }
  }
}

\newcommand{\ratess}[2]{\displaystyle\mathrel{\longrightleftarrows^{#1\mathstrut}_{#2}}}

\def\longrightarrowup{
  \mathop{
    \vcenter{
      \hbox{
	\ooalign{
	  \hbox{$\longrightarrow\joinrel$}\crcr
	}
      }
    }
  }
}

\newcommand{\rateup}[1]{\displaystyle\mathrel{\longrightarrowup^{#1\mathstrut}}}

\begin{abstract}
Understanding the causes and effects of spatial aggregation is one of the most fundamental problems in ecology. Aggregation is an emergent phenomenon arising from the interactions between the individuals of the population, able to sense only --at most-- local densities of their cohorts. Thus, taking into account the individual-level interactions and fluctuations is essential to reach a correct description of the population. Classic deterministic equations are suitable to describe some aspects of the population, but leave out features related to the stochasticity inherent to the discreteness of the individuals. Stochastic equations for the population do account for these fluctuation-generated effects by means of demographic noise terms but, owing to their complexity, they can be difficult (or, at times, impossible) to deal with. Even when they can be written in a simple form, they are still difficult to numerically integrate due to the presence of the ``square-root'' intrinsic noise. In this paper, we discuss a simple way to add the effect of demographic stochasticity to three classic, deterministic ecological examples where aggregation plays an important role. We study the resulting equations using a recently-introduced integration scheme especially devised to integrate numerically stochastic equations with demographic noise. Aimed at scrutinizing the ability of these stochastic examples to show aggregation, we find that the three systems not only show patchy configurations, but also undergo a phase transition belonging to the directed percolation universality class.
\end{abstract}

\section{Introduction}
The connection between different scales of observation is a central issue in many disciplines. The most suitable level of resolution to scrutinize a given system depends essentially on the type of questions to be answered. A description at the level of the individual components --in which all elementary interactions are taken into account-- contains all the information, but at the price of being, in general, intractable analytically or, in some cases, even computationally.  In the same way as in physics the Navier-Stokes equations are used to describe fluid dynamics in terms of coarse-grained fields (by-passing the description in terms of individual molecules), in ecology it is usual to resort to population-level dynamical equations encapsulating the most relevant features of groups composed of a large number of individuals. In this approach, individuals are replaced by ``fields'', which account for the density of organisms at specific points of space and time. Thus, deducing equations for those density fields starting from individual interactions, without losing relevant information, is an essential task.

Almost two decades ago, Durrett and Levin reviewed the standard (coarse-grained) modeling approaches in population ecology \cite{DL}. They discussed different levels of description (individual level and continuous equations, in both their spatially explicit and implicit versions) as well as their mutual interconnections. Each of these approaches is able to reproduce some features correctly, even if with limitations. All the continuous descriptions presented in \cite{DL} are deterministic, and therefore suppress the stochasticity inherent in the individual level. Even the ``hydrodynamic approach'', introduced in \cite{DL} and extended in \cite{CC}, which does account for the discreteness of the individuals, eventually leads to deterministic continuous equations at the population level, in which stochasticity is averaged away by assuming a Poissonian distribution for the number of individuals. 

In most cases, deterministic descriptions suffice to capture successfully the population-level phenomenology and are likely to be analytically tractable. Thus, they may provide a global understanding of the system phenomenology and allow us to quantify the effect of different factors.  However, in some other cases neglecting stochastic effects leads to the loss of relevant information. In particular, deterministic approaches fail to describe noise-induced effects, which can be crucial in certain situations especially for low population densities and low spatial dimensions \cite{Amit}. To focus our presentation, here we restrict ourselves to the role of noise in the problem of pattern formation in ecology. We shall discuss three different examples.

Based on the early work of Turing \cite{Turing} and others \cite{steele0}, Levin and Segel (and, independently, Okubo \cite{okubo}) introduced a model describing the dynamics of interdependent phytoplankton and zooplankton populations at a deterministic (highly coarse-grained) level \cite{LS}.  The corresponding differential equations develop characteristic (Turing) patterns of aggregation when the differential diffusion of the two species is large, while homogeneous stationary states are obtained otherwise \cite{LS}.  In particular, the region in the parameter space for which patterns are observed is relatively narrow as opposed to real planktonic populations, which typically appear in patchy distributions even at scales not traceable to physical forcing. Moreover, the assumption of passive diffusion for zooplankton cannot be biologically justified, since zooplankton can actively aggregate, and this is reflected in the fact that model predictions of phytoplankton being more patchily distributed than zooplankton are not borne out in empirical data.

Aimed at clarifying this paradoxical situation, Steele and Henderson considered a model very similar to the Levin-Segel Model (LSM), but including stochasticity (i.e. a Gaussian white noise) in a parameter value \cite{steele}. Such a noise term --mimicking intraspecific variability in either phytoplankton or zooplankton-- allows one to scrutinize whether ecological interactions suffice to explain patchiness, with no need to invoke external (environmental) variation or fluctuations, or active aggregation. Actually, the introduction of noise results in a widening of the region of patchiness in the parameter space and, therefore, serves as a possible explanation for the origin of pattern formation in some real-life examples \cite{steele}.  In the same spirit, Butler and Goldenfeld recently showed that, indeed, fluctuations can expand the pattern region in the LSM \cite{BG}. First, they showed that the naive method of adding by hand a simple white noise to the deterministic LSM is able to induce patterns of aggregation under less constrained condition. Then, they introduced an individual-level model representing the interactions between the two planktonic types, for which they performed a rigorous scaling-up by employing standard techniques from statistical physics and field theory \cite{BG,BG_new,GF}. This approach allowed them to derive a set of stochastic (Langevin) equations whose deterministic part coincides with that of the original LSM but where, additionally, there is a non-trivial stochastic part or noise (that includes non-trivial correlations, cross-correlations, and diffusive noise) \cite{BG,BG_new}.  These {\it demographic noise} terms are a direct consequence of the stochastic nature of the underlying birth and death processes. Their elegant analytical calculation reveals that demographic or ``intrinsic'' noise greatly enlarges the region of the parameter space where pattern formation occurs. Hence, even in the absence of either environmental or intraspecific variability, patterns appear much more generically than in the purely deterministic LSM as a mere consequence of demographic noise.

A second example in which intrinsic stochasticity due to discreteness plays a relevant role in pattern formation is the study of vegetation growth in semi-arid environments. Klausmeier introduced a continuous model consisting of two coupled differential equations representing the interactions between vegetation biomass and water \cite{chris}. In the presence of terrain slopes or soil inhomogeneities, the model develops Turing and/or disordered patterns \cite{chris,chris_model_PRL}. However, the model is not able to generate patterns in plain terrains, for which only homogeneous distributions of vegetation exist. Shnerb and collaborators utilized a hybrid approach to this problem \cite{shnerbJTB}. They considered the deterministic equations governing the vegetation model, but introduced an additional ``integration trick'' aimed at incorporating in some effective way the discrete nature of the underlying individuals/plants and, hence, to incorporate indirectly stochastic demographic effects. Thus, the question arises as to whether it is possible to account for these discreteness-induced effects purely at the coarser level of the population, that is, introducing stochasticity in the deterministic equations in a more systematic and controlled way.  Again, thanks to the methods and concepts developed in field theory \cite{GF}, we know that the answer is yes. As in the example of plankton above, it is possible to derive analytically a set of Langevin equations starting from the discrete model by using the same approach. Once more, however, the resulting set of equations is technically difficult to deal with.

A third example to be discussed is the formation of regular, almost crystal-like, patterns in bacterial colonies.  To this end, we use the simple ``Brownian bug model'', in which random walkers (bugs) diffuse, branch, and die at some rates. While deterministic approaches to this model lead to too-perfect orderings, incompatible with fundamental principles of physics (see below), the inclusion of stochasticity leads to much more realistic patterns, as well as accurate and precise predictions \cite{bugs1,bugs2,framos}.

Summing up, at the end of this discussion, we are left with the following dichotomy: either we have {\it i)} an individual-based description very detailed but lacking emphasis on large-scale features and, hence, handicapping the understanding of the emerging phenomenology, or {\it ii)} a deterministic global description in terms of density fields, which emphasizes large-scale properties but potentially misses important features, specially in low-dimensional systems and for low densities.  As described above, in some cases a third way exists: it is possible, starting from individual-based models, to derive analytically stochastic continuous (Langevin) equations. They usually are extensions to their deterministic counterparts that include additional demographic-noise terms. For instance, a recent example of the successful application of this systematic approach to the spatial Lotka-Volterra model can be found in \cite{tauber}. Unfortunately, such Langevin equations are usually difficult to treat through computational studies, leaving analytical approaches (which may or may not be feasible) as the only available options \cite{footnote}.

The essential problem one faces in trying to study computationally Langevin equations with demographic noise is that standard integration methods generate unrealistic negative density values \cite{neg_dickman}.  Demographic noises are proportional to the square-root of the density field and, owing to this, whenever the local values of the density become close to zero, the noise term is much larger in magnitude than any other term in the equation. This, combined with the random sign of the noise term, leads standard integration schemes  ineluctably to negative densities and, hence, to numerical instabilities.  However, a novel and powerful integration method has been developed to deal with demographic noise in an accurate and precise way \cite{DCM,moro}.  The method is a {\it split-step} algorithm in which the system is discretized in space, and in which temporal integration is implemented in two steps: (i) The noise term plus linear terms in the deterministic dynamics are treated in an exact way by sampling the conditional probability distribution coming out of the (exactly solvable) associated Fokker-Planck equation. By sampling such a distribution, an output is produced at each site. (ii) Then, the remaining deterministic terms are integrated using any standard scheme, choosing as initial conditions the output of the previous step at each site.  This scheme is able to avoid the difficulties associated with negative density values, and converges to the Langevin equation solution for discretization times that need not be as small as in the standard integration approaches (see \cite{DCM,moro} for further details).

In this paper, we apply the split-step integration scheme to Langevin equations describing the three examples discussed above. We focus on the (hardest-to-analyze) small-density limit, i.e.  close to extinction. To avoid further difficulties (stemming mostly from noise cross-correlations and conserved noise terms) we use a minimal description in which we take the deterministic equations for each case and add the simplest possible form of (biologically reasonable) demographic noise to each of them. This can be considered as either a stochastic extension of the deterministic models, or as a simplification of the more complex Langevin equations analytically derived from the corresponding individual-based model, in which higher-order irrelevant terms are omitted.  By using the split-step integration scheme, we scrutinize numerically the role of demographic noise in the generation of patterns in the stochastic versions of the examples presented above. Observe that, in all these examples, when extinction is reached (i.e. when the density of phytoplankton, vegetation or bacteria vanishes) the system remains trapped in such a state indefinitely: phytoplankton, vegetation, and bacteria do not arise spontaneously and, therefore, the empty state is an absorbing state. Thus, it is also interesting to investigate if the behavior of our Langevin equations belongs to any known universality class for systems with absorbing states \cite{hinrichsen,odor,GR,marro}.  We show that this minimal continuous stochastic approach keeps the relevant fluctuations present at the discrete level in the studied examples and, in consequence, is able to reproduce the most significant phenomenology and key features observed in the microscopic-level counterparts.

\section{Patterns in  Oceanic Plankton}
The Levin-Segel model (LSM) is defined by a set of two deterministic equations accounting for the basic interactions between a population of phytoplankton (autotrophic organisms) and zooplankton (heterotrophic organisms that feed on phytoplankton, i.e. grazers) \cite{okubo,LS}. If $\rho(\textbf{x},t)$ is the density of phytoplankton at position $x$ and time $t$, and $\phi(\textbf{x},t)$ the density of zooplankton at the same coordinates, the LSM may be written as:

\begin{equation}
\textnormal{LSM}\;\;\Longrightarrow\left\lbrace\vspace{0.25cm}
\begin{array}{rl}
\partial_{t}\rho(\textbf{x},t)&=a\rho+b\rho^{2}-w\rho\phi+D\nabla^{2}\rho
\\ \partial_{t}\phi(\textbf{x},t)&=w_{2}\rho\phi-\lambda\phi^{2}+D_{2}\nabla^{2}\phi,
\end{array}
\right.
\label{LS_det}
\end{equation}
\noindent
where $a$, $b$, $w$, $D$, $w_{2}$, $\lambda$ and $D_{2}$ are constants \cite{note0}. The first two terms on the right hand side of the first equation describe the growth and replication of phytoplankton cells, the third one the mortality due to zooplankton consumption, and the fourth one the diffusion of cells. Similarly, the first and last terms on the r.h.s. of the second equation describe the growth of the zooplankton population due to grazing and effective diffusion, respectively, while the second one is a saturation term imposing a certain carrying capacity.

Together with the trivial extinction state, $\rho(x,t)=\phi(x,t)=0$, and the (biologically implausible) state in which $\phi$ vanishes and $\rho$ diverges, Eqs.(\ref{LS_det}) have a spatially uniform stable equilibrium in which  $\rho=a\lambda/(w w_{2}-b\lambda)$ and $\phi=a w_{2}/(w w_{2}-b\lambda)$, provided $w w_{2} > b\lambda$ and $w_{2} > b$. In the usual way, the separation of scales between diffusion constants allows for Turing patterns to emerge \cite{Turing}; in particular, when the ratio $D_{2}/D$ is larger than a certain constant $K_{cr}$ (which is a combination of some parameters \cite{LS}), the homogeneous state becomes unstable, entailing pattern formation.  Thus, the line $D_{2}/D=K_{cr}>1$ separates the zones of the parameter space where it is possible to generate Turing patterns from those where the homogeneous state is stable (see blue curve of Fig.1 in \cite{BG}).

As mentioned above, an individual-level version of that model was introduced by Butler and Goldenfeld \cite{BG}. Based on the interactions between phytoplankton ($P$)and zooplankton ($Z$) proposed by the LSM, it explicitly writes reaction equations for the different sources of reproduction and mortality:

\begin{equation}
\begin{array}{rl}
P\rateup{a} PP \;\;\;\;\; & PP\rateup{b/V} PPP\\
PH\rateup{ w /V} HH \;\;\;\;\; & HH\rateup{\kappa/V} H
\end{array}
\label{LS_IBM}
\end{equation}

\noindent
(where $V$ is the volume of a well-mixed patch where these reactions are considered, and $w=w_{2}$ is assumed), together with the passive diffusion of the agents. From the proposed set of individual interactions, by using standard field-theory techniques \cite{GF,vKampen}, a set of coupled Langevin equations for the density fields was deduced.  Such equations coincide at the deterministic level with Eq.(\ref{LS_det}), but they include additional noise terms (see Eq.(15) in \cite{BG}).  As shown analytically by the authors, this stochasticity greatly enlarges --owing to a resonant amplification of fluctuations \cite{SAF}-- the region in which patterns emerge, leaving only a small region of parameter space with homogeneous stable solutions. 

The actual complex form of the noise in these Langevin equations makes it difficult (if not impossible) to verify numerically the analytical result at the mesoscopic level. Of course, it is always possible to perform direct simulations at the microscopic level, but here we are interested in the mesoscopic --Langevin-- description. In order to incorporate to the relevant individual-level fluctuations to the set of discrete equations in Eqs.(\ref{LS_det}), we add demographic-noise terms to each of the differential equations in the simplest possible form compatible with biological constraints:

\begin{equation}
\textnormal{S-LSM}\;\;\Longrightarrow\left\lbrace\vspace{0.25cm}
\begin{array}{rl}
\partial_{t}\rho(\textbf{x},t)&=a\rho+b\rho^{2}-w\rho\phi+D\nabla^{2}\rho+\sigma\sqrt{\rho}\;\eta(\textbf{x},t)
\\ \partial_{t}\phi(\textbf{x},t)&=w_{2}\rho\phi-\lambda\phi^{2}+D_{2}\nabla^{2}\phi+\sigma_{2}\sqrt{\rho\phi}\;\eta_{2}(\textbf{x},t),
\end{array}
\right.
\label{LS_Lang1}
\end{equation}

\noindent
where $\sigma$ and $\sigma_{2}$ are constants, and $\eta$ and $\eta_{2}$ are delta-correlated Gaussian noise (i.e. white noise) terms \cite{note3}. Heuristically, one can argue that --as it is usually the case in particle systems, and as a direct consequence of the central limit theorem-- the noise amplitude has to be  proportional to the square-root of the involved densities. Observe that, in the case of zooplankton, reproduction and mortality are related to the presence of phytoplankton; therefore, the variance of the new noise term must be proportional to both zooplankton and phytoplankton densities; for phytoplankton, the leading order of approximation leaves the variance of the noise proportional to the phytoplankton density. Alternatively, it is not difficult to show that these terms, present in the detailed derivation in \cite{BG}, are the leading terms in that approach. Higher-order corrections are irrelevant in the renormalization group sense and noise cross-correlations, which might not be irrelevant, are anyhow perturbatively generated. The advantage of our simplification is that, now, the resulting set of Langevin equations can be numerically studied by direct application of the split-step integration scheme for Langevin equations with square-root noise \cite{DCM,moro}.

\begin{figure}[ht]
\begin{center}\includegraphics[width=0.6\textwidth]{phase_diagram2.eps}\\
\includegraphics[width=0.5\textwidth]{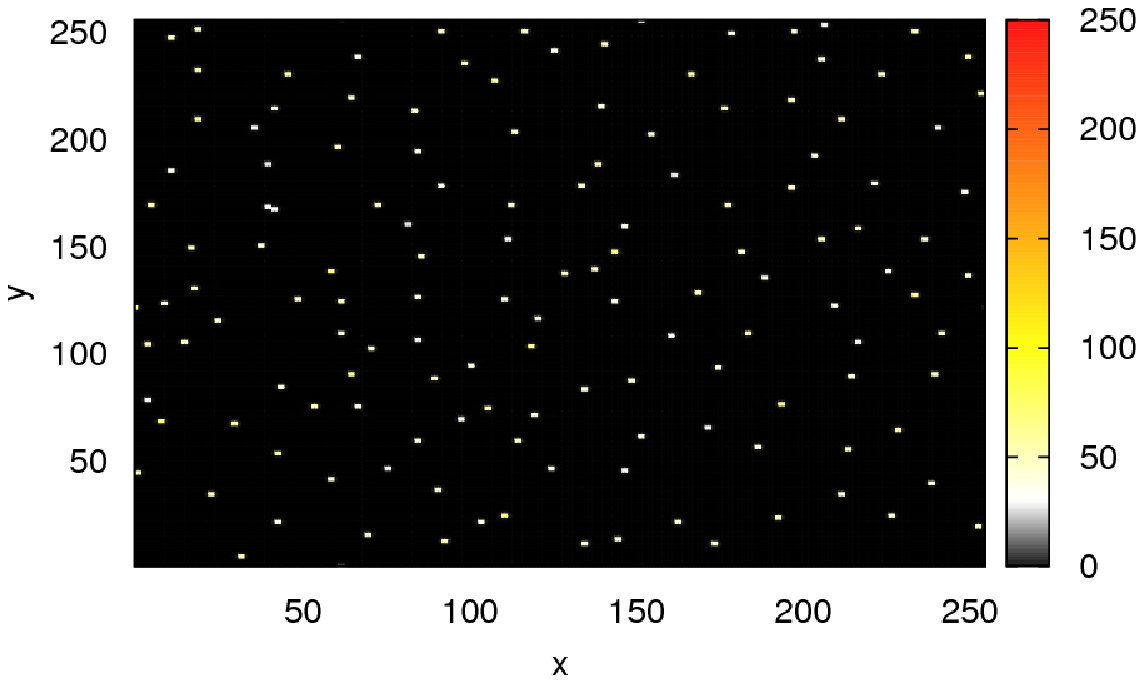}\hspace*{0.2cm}
\includegraphics[width=0.5\textwidth]{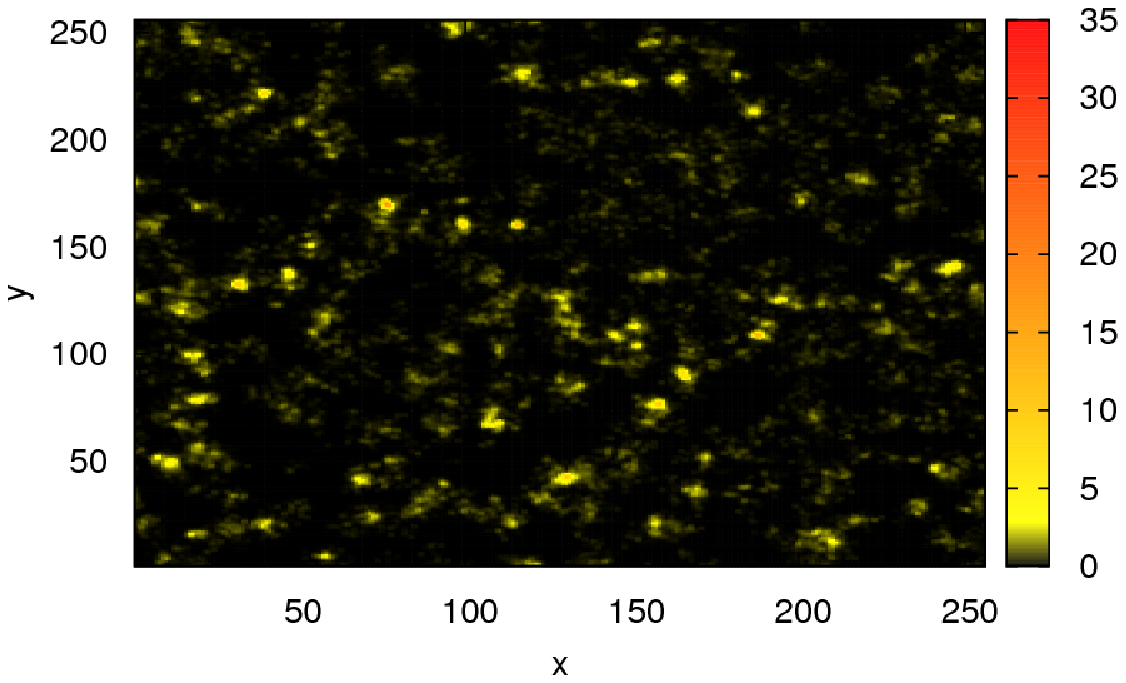}\end{center}
\caption{\footnotesize{Up: Generic phase diagram of the stochastic version of the LSM. Down: Representative examples of the disordered patterns formed by the phytoplankton population in the supercritical phase.}}
\label{LSM_phase_diagram} 
\end{figure}

We use the integration scheme in a $2$-dimensional lattice of lateral size $L$. We study the change in time of the spatially-averaged density of both fields, $\overline{\rho}$ and $\overline{\phi}$, starting from a homogeneous state, and determine how that behavior is altered by changing the control parameter $a$ while fixing the rest of parameters in Eqs.(\ref{LS_Lang1}).  As happens with the original LSM, we find two phases: an active phase, where both phytoplankton and zooplankton reach a stationary non-trivial state, and an absorbing phase, where the primary extinction of phytoplankton leads unavoidably to zooplankton extinction.  The integration algorithm permits us to reproduce generically patterns almost all along the active phase, allowing for a numerical verification, at the mesoscopic level, of the analytical findings in \cite{BG} (see Fig.\ref{LSM_phase_diagram}).

Note that the lack of a saturation term for phytoplankton in Eqs.(\ref{LS_Lang1}) could lead to the erroneous conclusion that its averaged density could grow boundlessly in the supercritical phase (i.e. no stationary state would be possible). Interestingly, it is the coupling between fields in the zooplankton equation that keeps the phytoplankton field bounded: when phytoplankton density grows, the positive coupling in the $\phi$ equation entails a consequent zooplankton density growth which, in turn, has a bounding effect on phytoplankton density. Thus, zooplankton acts as an inhibitor and stabilizes the phytoplankton population. 

\subsection*{Critical properties and universality}

\begin{figure}[ht]
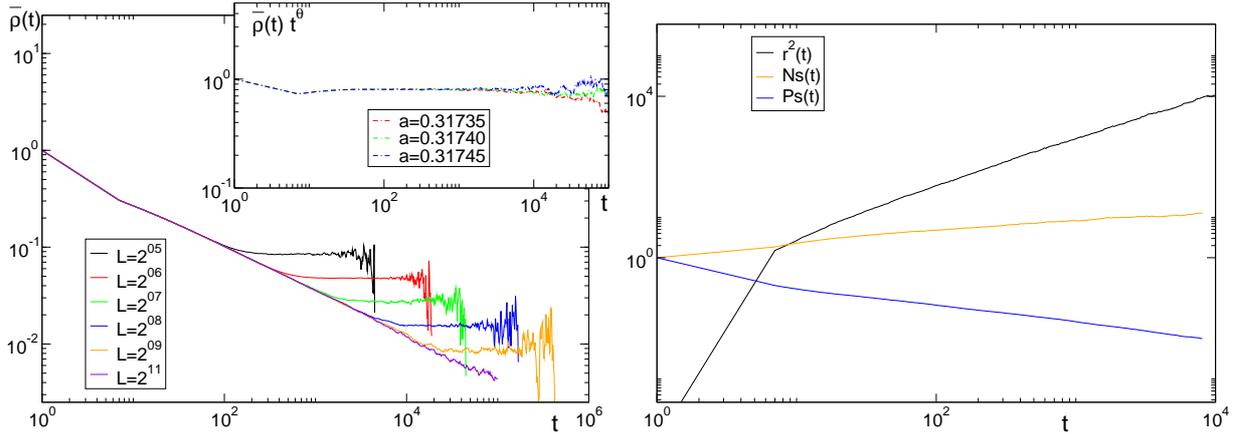

\includegraphics[width=0.5\textwidth]{theta_all_740_w_inset_cpoint.eps}\hspace*{0.1cm}
\includegraphics[width=0.5\textwidth]{spreading_740.eps}
\caption{\footnotesize{Dynamic behavior of the S-LSM with parameters $b=0$, $w=w_{2}=1$, $D=D_{2}=0.25$, $\lambda=1$, $\sigma=\sigma_{2}=\sqrt{2}$ and $dt=0.1$. Left: Temporal decay (in  MonteCarlo steps) of the spatial average of phytoplankton density, starting from a homogeneous distribution, for different system sizes. Inset: Resulting curve of multiplying $\overline{\rho}(t)$ by $t^{\theta}$, using a system of linear size $L=2^{11}$ (i.e. $4,194,304$ sites); the horizontal curve corresponds to the critical point. Right: Mean quadratic radius (upper curve), number of sites with non-zero density and survival probability (lower curve) for the spreading of a localized seed of phytoplankton and zooplankton in an otherwise empty system.}}
\label{LS1_temp} 
\end{figure}

To study universality issues at the transition point, and in deference to the simplest biology (without active aggregation), we relax the assumption that $D_{2}\gg D$ and, actually, set $D=D_{2}$ to a small value --see caption of Fig.\ref{LS1_temp}. This enhances the convergence to the asymptotic state while avoiding the presence of ``Turing-like'' effects (as $D/D_{2}=1$). As the condition $b\neq0$ in Eqs.(\ref{LS_det}) is essential for both the original \cite{LS} and extended \cite{BG} versions of the model to show a non-homogeneous solution, we also impose $b=0$ in order to challenge the ability of our new description to show patterns.

As can be seen in the left panel of Fig.\ref{LS1_temp} (inset), starting from a homogeneous initial condition, for values of $a$ below $a_c=0.31740(5)$ and large system sizes there is an initial power-law decay of the phytoplankton population density which eventually goes extinct in a finite time. For values of $a$ above such value, the population eventually reaches a stationary state with a non-zero density for both species. At $a=a_{c}$, the system undergoes a second-order phase transition characterized by the scale-invariant decay of the density following a power law of exponent $\theta=0.49(5)$. 

\begin{figure}[ht]
\begin{center}\includegraphics[width=0.6\textwidth]{static_all_740.eps}\\
\includegraphics[width=0.5\textwidth]{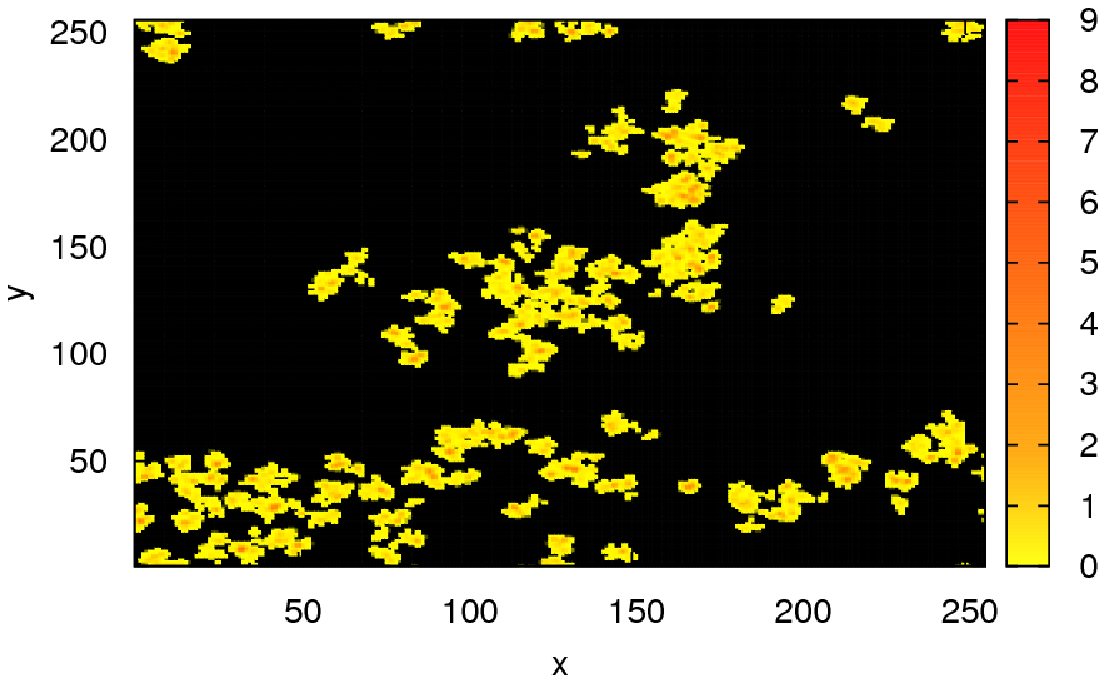}\hspace*{0.2cm}
\includegraphics[width=0.5\textwidth]{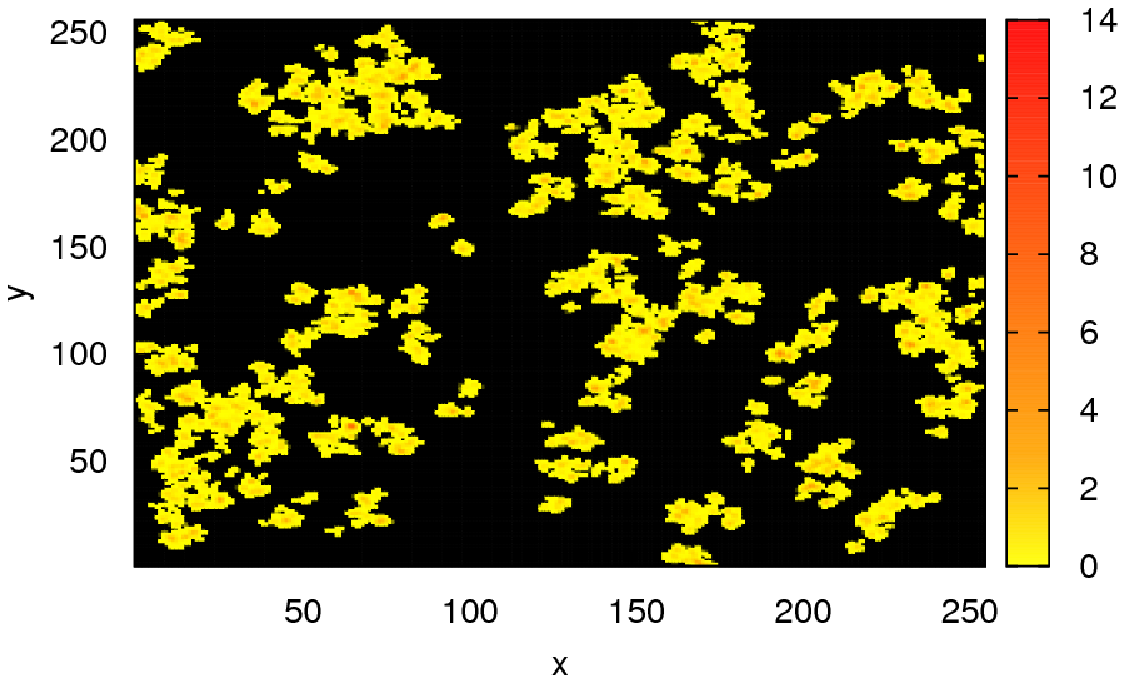}\end{center}
\caption{\footnotesize{Up: Finite-size scaling of the stationary values depicted in Fig.\ref{LS1_temp} (left panel) and the survival time for homogeneous experiments when using a threshold for the survival  probability of $0.1$. Down: Typical snapshot of the phytoplankton density in the supercritical phase.}}
\label{LS1_stat} 
\end{figure}

As a consequence of the existence of a critical point, one expects scale-invariant behavior of other  quantities.  {\it i)} Performing simulations starting from an initial small density for both fields, localized at a single point, we can study how the population spreads over an otherwise empty system. At the critical point, the mean quadratic radius of the population (mean square distance of the population border to the original ``seed''), $r^{2}$, the number of lattice sites with a non-zero density of any of the two species, $N_{s}$, and the survival probability of the population, $P_{s}$, change with time following a power law of exponents $z_{spr}=1.13(2)$, $\eta=0.23(2)$ and $\delta=0.42(2)$, respectively (see right panel of Fig.\ref{LS1_temp}). {\it ii)} The stationary density of the smaller sizes (see left panel of Fig.\ref{LS1_temp}) on the linear system size $L$, follows a power law described by $\rho_{st}\sim L^{\beta/\nu_{\perp}}$. This is also the case of the survival time of experiments started from homogeneous conditions, $t_{sur}$, which depends on $L$ following a power law of exponent $\nu_{\parallel}/\nu_{\perp}$ (see table 1 for actual values, and left panel of Fig.\ref{LS1_stat}). 

\begin{table}[ht]
\begin{center}
\begin{tabular}{|c||c|c|c|c|c|c|c|}
\hline
&$\theta$&$\delta$&$\eta$&$z_{spr}$&$\beta/\nu_{\perp}$&$\nu_{\parallel}/\nu_{\perp}$\\
\hline
DP \cite{DP,DP2}&$0.4505(10)$&$0.4505(10)$&$0.2295(10)$&$1.1325(10)$&$0.795(4)$&$1.766(2)$\\
\hline
S-LSM&$0.49(5)$&$0.42(2)$&$0.23(2)$&$1.13(2)$&$0.83(5)$&$1.73(5)$\\
\hline
S-Shnerb&$0.47(5)$&$0.48(2)$&$0.23(2)$&$1.16(2)$&$0.79(5)$&$1.75(5)$\\
\hline
\end{tabular}
\end{center}
\label{table}
\caption{Critical exponents obtained with the simple stochastic versions of the deterministic models analyzed in the text, plus the corresponding values for the directed percolation (DP) universality class. The exponents for the ``Brownian Bug`` model, in agreement with the DP universality class, can be found in \cite{BB_exp}.}
\end{table}

The right panel of Fig.\ref{LS1_stat} is an illustrative example of the spatial distribution of the phytoplankton population in the supercritical stationary states. As we can see, phytoplankton form disordered patterns with patches of different sizes with non-zero density. 

In summary, the set of equations Eqs.(\ref{LS_Lang1}) with $b=0$ and $D/D_{2}=1$ is not only able to describe patchiness in its spatial distribution (a common feature or real phytoplankton-zooplankton populations \cite{LS,Simon2}), but also shows a richer, non-trivial phenomenology not observed in the original deterministic formulation, which stems from stochasticity. Careful inspection reveals that the derived set of critical exponents (see table 1) corresponds to the universality class of directed percolation, which controls many transitions into absorbing states \cite{hinrichsen,odor,GR,marro}. The reasons why this model, as well as the rest of models discussed in the paper, belongs to the Directed Percolation universality class are discussed in the Appendix. Analogously, our approach also permits us to study a region of parameters in which either supercritical DP clusters or other non-trivial patterns emerge (see Fig.\ref{LSM_phase_diagram}), and verify that this region is indeed largely enhanced with respect to the deterministic predictions.

\section{Patterns in Vegetation in Semi-Arid Environments}

Aimed at describing the patterns of vegetation observed in semi-arid zones, Klausmeier introduced a model that keeps track of the interactions between vegetation biomass and water in such ecosystems \cite{chris}. This model, as well as an extension to it including a more complex water diffusion term \cite{chris_model_PRL}, leads to pattern formation only in the ``Turing limit'' (i.e. very different diffusion rates for the vegetation and water density fields), or in inhomogeneous media.

Shortly afterward, Shnerb et al. revisited the problem in a series of papers \cite{shnerbJTB,shnerbPRL,shnerbPRE,shnerbEPL}, introducing a simplified, combined version of the two previously mentioned models with the intention of reproducing the disordered patterns frequently observed in semi-arid plains. If $\rho$ represents the biomass density, and $\phi$ water density, the equations read:

\begin{equation}
\textnormal{Shnerb}\;\;\Longrightarrow\left\lbrace\vspace{0.25cm}
\begin{array}{rl}
\partial_{t}\rho(\textbf{x},t)&=-a\rho+w\rho\phi+D\nabla^{2}\rho \\
\partial_{t}\phi(\textbf{x},t)&=R-w_{2}\rho\phi-\kappa\phi+D_{2}\nabla^{2}\phi- \textbf{v} . \nabla\phi, 
\end{array}
\right.
\label{shnerb_det}
\end{equation}

\noindent
where $a$, $R$, $w$, $w_{2}$, $\kappa$, and $D$ are constant parameters, and $\textbf{v}$ is a constant velocity vector. In this model, vegetation (first equation) grows in the presence of water, there is an effective diffusion due to, e.g. seed dispersal, and there is also competition for water, here represented by the first term on the r.h.s. On the other hand, water density (second equation) increases due to precipitation ($R$) and decreases due to consumption and evaporation (second and third term on the r.h.s.); it can also flow from high to low places, as represented by the last term of the r.h.s.

Eqs.(\ref{shnerb_det}) exhibit two different homogeneous equilibria: a trivial state where vegetation goes extinct and water reaches the stationary value imposed by precipitation ($\rho=0$, $\phi=R$), and a nontrivial one where both reach a stationary state given by ($\rho=(R-a)/(a w_{2})$, $\phi=a$). In the absence of cross-diffusion terms like the one proposed in \cite{chris_model_PRL} or anisotropies such as the one imposed by $v\ne0$, the system lacks unstable homogeneous solutions that can be perturbed in order to obtain the desired patterns.
 
Shnerb and coauthors realized that Eqs.(\ref{shnerb_det}) are able to show realistic disordered patterns when a seasonal removal of biomass below a certain threshold is introduced \cite{shnerbJTB,shnerbPRL}, and/or an additional selective mortality that depends on the neighboring biomass \cite{shnerbJTB}. Both mechanisms account in some effective way for stochastic effects at low densities, although they somehow mix different scales of description.

We follow now the steps of the previous section to reproduce disordered patterns in a systematic way without resorting to integration prescriptions \cite{note4}.  In analogy with the previous section, we introduce a simple demographic noise term in the equation for the vegetation biomass density. As in the case above, we set $v=0$ (i.e. no anisotropy or inhomogeneous soil) and $D/D_{2}=1$. The resulting equations for the stochastic version of the Shnerb model (S-Shnerb) are:

\begin{equation}
\textnormal{S-Shnerb}\;\;\Longrightarrow\left\lbrace\vspace{0.25cm}
\begin{array}{rl}
\partial_{t}\rho(\textbf{x},t)&=-a\rho+w\rho\phi+D\nabla^{2}\rho+\sigma\sqrt{\rho}\eta(\textbf{x},t) \\
\partial_{t}\phi(\textbf{x},t)&=R-w_{2}\rho\phi-\kappa\phi+D_{2}\nabla^{2}\phi,
\end{array}
\right.
\label{shnerb_Lang}
\end{equation}

\noindent
where the terms and constants are as explained for Eq.(\ref{shnerb_det}), and the new term is identical to the one introduced for phytoplankton in the previous example.

Using again the split-step integration scheme, and choosing the precipitation density $R$  as tuning parameter, we observe that, for $R<R_{c}=0.40835(5)$, the spatially-averaged density of vegetation biomass decays until the population becomes extinct, while for $R>R_{c}$, it reaches a non-trivial stationary state. The system undergoes a phase transition between these two states at $R=R_{c}$, where $\overline{\rho}$ decays asymptotically toward extinction following a power-law. On the other hand, the averaged water field, $\overline{\phi}$, eventually reaches a stationary value in any of the three situations. For $R<R_{c}$, $\overline{\phi}$ grows with $R$, while for $R>R_{c}$, that value decreases as $R$ (and $\rho_{st}$) increases.

Once more, the biological processes involved shed light on how the stationary state is achieved in the absence of a saturation term in Eq.(\ref{shnerb_Lang}). As described above, when the vegetation population goes  extinct, larger values of the precipitation constant $R$ lead to larger values of the average water density $\overline{\phi}$. On the other hand, there is a change in trend when vegetation survives. For values of $R$ into the supercritical phase, the larger the precipitation parameter is, the better the conditions for vegetation growth are. In consequence, the average biomass $\overline{\rho}$ reaches a larger value at the stationary state, and the improved consumption influences the average value of the water density, which decreases with the increase of $R$. Therefore, increasing $R$ entails a decrease in the average of linear terms in the $\rho$ equation, which keeps bounded the biomass field. The competition for the available resource (water) acts as an inhibitor, which allows the vegetation density to reach a well-defined stationary state, acting as an effective carrying-capacity term.

\begin{figure}[ht]
\begin{center}\includegraphics[width=0.6\textwidth]{spreading_835_w_inset.eps}\\
\includegraphics[width=0.5\textwidth]{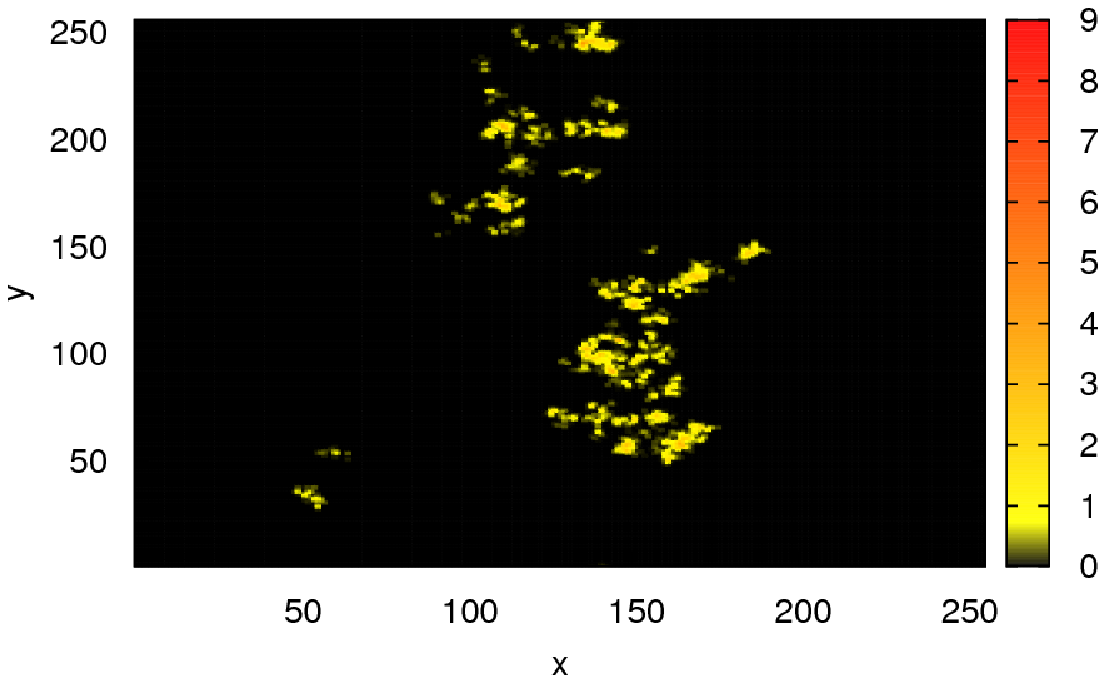}\hspace*{0.2cm}
\includegraphics[width=0.5\textwidth]{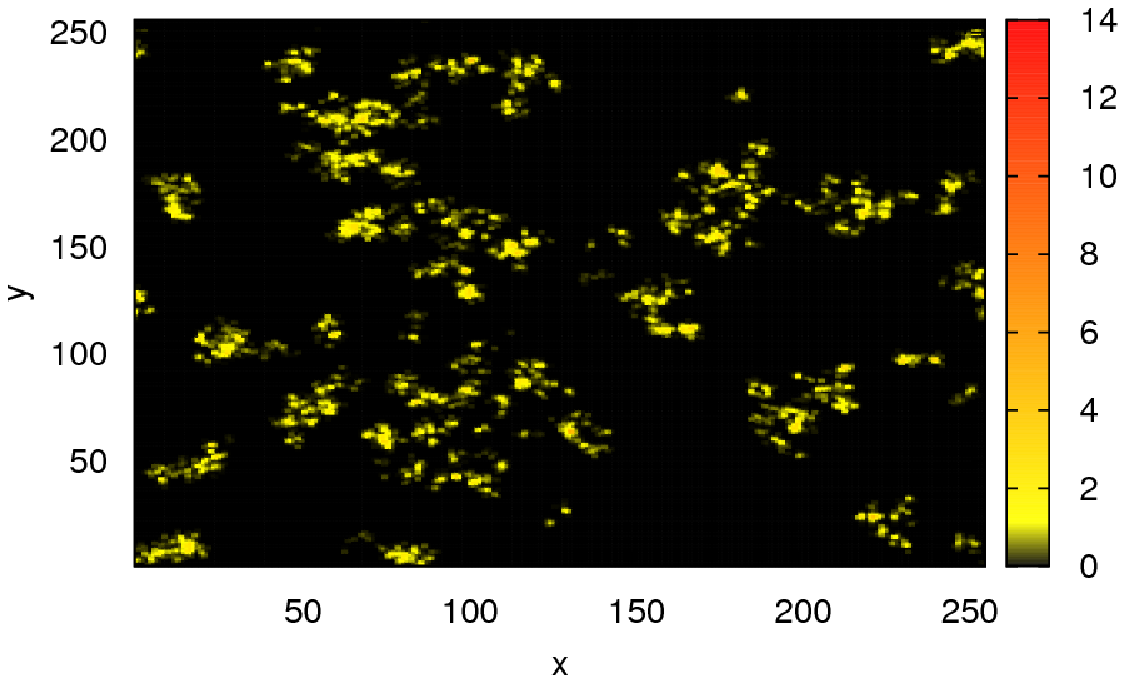}\end{center}
\caption{\footnotesize{Behavior of Eqs.(\ref{shnerb_Lang}) with $a=0.2$, $w=1$, $w_{2}=1.2$, $\kappa=1$, $D=D_{2}=0.25$ and $\sigma=\sqrt{2}$. Up: Spreading observables (main panel) and stationary observables (inset) for $R=R_{c}=0.40835$. Down: Typical snapshot of the vegetation density in the supercritical phase.}}
\label{shnerb_all} 
\end{figure}

Analyzing both the dynamic and stationary behaviors of Eqs.(\ref{shnerb_Lang}), we obtain the curves and exponents shown in Fig \ref{shnerb_all} and table 1. As in the previous case, the existence of a critical point entails scale invariance that is translated into power-law observables and disordered clusters of vegetation of any size. For $R=R_{c}$ and above, it is possible to find the desired disordered patterns with these equations (see right panel of Fig.\ref{shnerb_all}). As in the example above, the measured critical exponents agree (within error bars) with those of the DP class (see table 1 and Appendix).

\section{Brownian bugs}

As a third and final example, we present an individual-based model: the ``interacting Brownian bug'' (IBB) model \cite{bugs1,bugs2}.  It consists of branching-annihilating Brownian particles (bugs, bacteria, etc) interacting with each other within a finite distance, $l$ \cite{bugs1}.  Particles move off-lattice in a $d$-dimensional space, and their dynamics is such that they can diffuse at rate $1$, performing Gaussian random jumps of variance $2D$; disappear spontaneously, at some rate $\beta_0$;  or branch, creating an offspring at their location with a density-dependent rate $\lambda$ modeling competition: 

\begin{equation}
 \lambda(j)= max  \{ 0, \lambda_0- {N_l(j)}/{N_{sat}} \},
\label{raterep}
\end{equation}
where $j$ is the particle label, $\lambda_0$ (reproduction rate in isolation) and $N_{sat}$ (saturation number) are fixed parameters, and $N_l(j)$ stands for the number of particles within a radius $l$ from $j$.  The control parameter is given by $\mu=\lambda_0-\beta_0$: the system is active above some $\mu_c$ and absorbing below. In the active phase, owing to the density-dependent dynamical rules, particles group together forming clusters with a characteristic size. Well inside the active phase, when these clusters start filling the available space, they {\it self-organize} in spatial patterns with remarkable hexagonal order.

The IBB model can be cast into a continuous stochastic equation \cite{bugs1,framos}.  Indeed, by applying the same standard techniques as above, a Langevin equation for the density $\rho$ of bugs can be derived \cite{framos}:

\begin{equation}
\dfrac{\partial \rho(\textbf{x},t)}{\partial t} = \mu 
\rho+D \nabla^{2} \rho
\label{langevin_bugs}
%\\ 
%\nonumber
-\frac{\rho}{N_s}\int_{\vert \textbf{x}-\textbf{y}
\vert < R} d \textbf{y} \rho(\textbf{y},t)+\sigma
\sqrt{\rho}\eta(\textbf{x},t),
\end{equation}

\noindent
where the noise amplitude $\sigma$ is a function of the microscopic parameters, $\eta(\textbf{x},t)$ is a normalized Gaussian white noise, and higher order terms and cross-correlations (irrelevant in the renormalization group sense) have been neglected. Remarkably, the deterministic part of Eq.(\ref{langevin_bugs}), including a non-local saturation term, is a particular case of the equation proposed by Fuentes {\it et al.}  \cite{Fuentes} as a model for competition-induced pattern formation, in which the non-linear term is:

\begin{equation}
- \rho(\textbf{x},t) \int d \textbf{y} F(\textbf{x}-\textbf{y})
\rho(\textbf{y},t)
\label{fuentes}
\end{equation}

\noindent
and $F$ is a generic kernel or influence function (which becomes a step function in Eq.(\ref{langevin_bugs})).

\begin{figure}[ht]
\center{\includegraphics[width=0.6\textwidth]{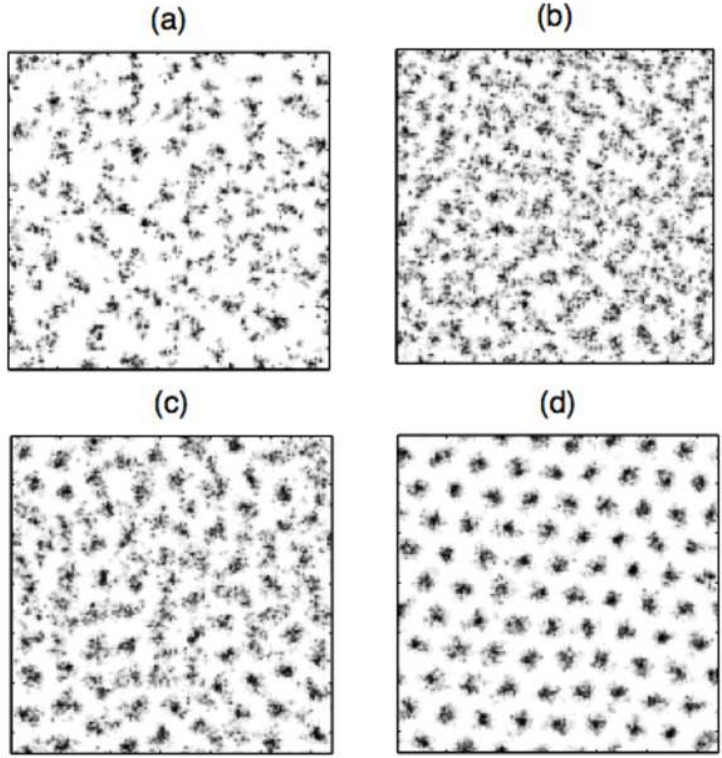}}
\caption{\footnotesize{Different patterns for different parameter values as obtained from a numerical integration of Eq. (\ref{langevin_bugs}). The amplitude of the noise decreases monotonously from (a) to (d).}}
\label{patterns}
\end{figure}

A numerical integration of Eq.(\ref{langevin_bugs}) relying on the split-step scheme has been recently performed \cite{framos}. It reproduces very accurately the phenomenology of the microscopic IBB model described above: there is a critical point in the DP class separating the patterned active phase from the absorbing phase. Two important observations are in order: {\it i)} As carefully discussed in \cite{framos} and illustrated in Fig.\ref{patterns} the resulting ordering is not ``perfect'' as would be the case for the corresponding deterministic equations. The presence of demographic noise induces ``defects'' and the resulting distribution of clusters is not a perfect regular crystal-like structure (which is actually precluded in two dimensional equilibrium systems by fundamental physics principles, i.e. the Mermin-Wagner theorem).  In fact, as shown in \cite{framos} the spatial structure (including defects, as well as their dynamics and statistics) is perfectly described by the theory of (equilibrium) two-dimensional melting. {\it ii)} As shown in \cite{Pigolotti} the case in which the kernel is Gaussian does not lead generically to clustering but rather to homogeneous solutions. However, it has been recently observed that, by introducing demographic noise, robust patterns similar to those reported above emerge. These are purely noise-induced patterns \cite{Simon_new,kernel}, that can be reproduced by our approach.

\section{Conclusions}

Connecting scales of observation is an important and delicate task. Important features of the individual level can be lost during the scaling up to the level of the population unless a rigorous approach is followed. In cases where a careful deduction of coarse-grained equations can be performed, these can be difficult to deal with. A trade-off between realism and complexity is to be resolved attending to the desired features to be reproduced by the continuous equations.

Simple forms of demographic noise, added to standard deterministic equations for different problems involving pattern formation, suffice to provide a much more precise description of the underlying ``microscopic'' dynamics.  We have considered three different examples: plankton dynamics, a model for vegetation growth in semiarid environments, and a model for interacting and diffusing particles. In all these cases, by adding demographic noise --even in its simplest form-- the regions in the parameter space where patterns appear are greatly enlarged.  Intrinsic noise also introduces a non-trivial phase transition separating the survival state from the absorbing or extinction one in all the reported examples.

This conclusion is backed by computational studies of the resulting stochastic/Langevin equation for each of the examples. Numerical integration of Langevin equations with demographic (``square-root'') noise is feasible owing to a recently proposed split-step integration scheme, avoiding the otherwise unavoidable integration instability associated with this kind of noise.

By using this split-step scheme, we have observed patchy distributions in regions for which the corresponding deterministic approach would only lead to homogeneous steady states. We have located a non-trivial phase transition separating the patchy active phase from the extinction/absorbing phase. Moreover, we have measured the critical exponents associated with such phase transition, finding that the universality class of the investigated examples is directed percolation. This universality class is paradigmatic of systems with absorbing states in the absence of other relevant symmetries.  In the two first discussed examples, we have seen that the coupling of activity to a non-conserved, diffusing field does not alter the universality class. Moreover, the absence of an explicit saturation term for activity results to be irrelevant here, as the (inhibitory) ecological interactions between the fields suffice to maintain the density of the interacting species bounded into finite intervals, allowing for a well-defined stationary state (either coexistence or extinction). 

In summary, properly derived Langevin equations --constructed either as an extension of deterministic models with additional demographic noise terms or as simplifications of more formally derived complicated stochastic equations-- provide a highly valuable tool for the analysis of population-level ecological properties. This is particularly true as long as numerical studies of such Langevin equations (a direct way to explore their phenomenology) are feasible. Simple Langevin equations combined with the split-step integration scheme provide us with a powerful tool to scrutinize population-level features keeping the relevant effects stemming from the underlying discrete nature of individuals. 

\section*{Acknowledgments}
We gratefully acknowledge support from the Cooperative Institute for Climate Science (CICS) of Princeton University and the National Oceanographic and Atmospheric Administration's (NOAA) Geophysical Fluid Dynamics Laboratory (GFDL), the National Science Foundation (NSF) under grant OCE-10046001, the Spanish MICINN-FEDER under project FIS2009-08451, and from Junta de Andaluc{\'\i}a (Proyecto de Excelencia P09FQM-4682). MAM thanks I. Dornic, H. Chat\' e and, C. L\'opez, and F. Ramos, for a long term collaboration on some of the issues presented here.

\section*{Appendix}

We have numerically shown that the critical exponents measured in all of the examples discussed in this paper are directed-percolation like. Here we justify that result by analyzing their corresponding Langevin equations Eqs.(\ref{LS_Lang1}), Eqs.(\ref{shnerb_Lang})), and Eq. (\ref{langevin_bugs}).

As conjectured years ago by Janssen and Grassberger \cite{conjecture} all systems exhibiting a phase transition into a unique absorbing state, with a single-component order parameter, and no extra symmetry or conservation law, belong to the same universality class whose most distinguished representative is directed percolation (DP).  In a field theoretical description this universality class is represented by the Reggeon Field theory (RFT), \cite{conjecture} which in terms of a Langevin equation reads:

\begin{equation}
{\displaystyle{\partial \rho (x,t) \over \partial t}} =  \nabla^2 \rho(x,t) +
 a \rho(x,t)
-b \rho^2(x,t) +
 \sqrt{\rho(x,t)} \eta(x,t)  ~~,
\label{RFT}
\end{equation}

\noindent
where $\rho(x,t)$ is the density field at position $x$ and time $t$, $a$ and $b$ are parameters, and $\eta(x,t)$ is a delta-correlated Gaussian noise, $<\eta(x,t) \eta(x',t')>= D \delta(x-x') \delta(t-t')$. 

The previous conjecture was confirmed in a large number of computer simulations, series expansion analysis, field theoretical studies, etc. Indeed, the DP universality class has proved to be extremely robust against the modification of many details in the microscopic models. The conjecture of universality has been extended systems with an infinite number of absorbing states \cite{Noi}. Also, Grinstein {\it et al.} extended the conjecture to the case of multicomponent systems \cite{GG}. Considering, for the sake of simplicity a two-component system with absorbing states, it can generically be described by a set of Langevin equations whose linearized dynamics (ignoring the Laplacian terms and the noise) is generically:

\begin{equation}
\begin{array}{rl}
\partial_{t}\rho_1(x,t)&=a_{1,1}\rho_1+ a_{1,2}\rho_2
\\ \partial_{t}\rho_2(x,t)&= a_{2,1}\rho_1 + a_{2,2}\rho_2
\end{array}
\label{Lang2}
\end{equation}

\noindent
and diagonalizing the matrix of the linear coefficients:

\begin{equation}
\begin{array}{rl}
\partial_{t}\xi_1(x,t)&=\lambda_1\xi_1 \\ \partial_{t}\xi_2(x,t)&=
\lambda_{2}\xi_2
\end{array}
\label{Lang3}
\end{equation}

\noindent
where $\lambda_1$ and $\lambda_2$ are the associated eigenvalues and $\xi_1$ and $\xi_2$ are the corresponding eigenvectors. The existence of the absorbing state implies that the two eigenvalues are negative. At the critical point one of the two eigenvalues, say $\lambda_1$ vanishes.  If the other one does not, it remains negative, and then fluctuations in $\xi_2$ continue to decay exponentially with time. This implies that $\xi_2$ does not experience critical fluctuations in the vicinity of the transition, and so can be integrated out of the problem. Hence, asymptotically, the set of Langevin equations can be reduced to single relevant equation which is precisely Eq.(\ref{RFT}), ensuing DP behavior.

On the other hand, if the secondary field is conserved (which, in particular, enforces $a_{2,1}$ and $a_{2,2}$ to vanish) or there is a hidden symmetry imposing $\lambda_1$ and $\lambda_2$ to vanish at the same point, this conclusion can break down. In both of these cases, at the system critical point, the two fields are simultaneously critical (or ``massless'' or ''gapless'' using the field theory jargon).

Systems in which the density field is coupled to a secondary {\it conserved} field indeed exhibit non-DP critical behavior.  This is the case of, for instance, {\it i)} systems related to self-organized criticality in which the activity field is coupled to a {\it conserved non-diffusive} background field \cite{FES} and {\it ii)} systems in which the secondary field is conserved but {\it diffusive} \cite{KSS,WOH}. 

Let us now return to the three models discussed in this paper.

\begin{itemize}
\item[{\it a)}] Let us first scrutinize the simpler case of vegetation patterns as described by Eqs.(\ref{shnerb_Lang}). At the critical point the expectation value of $\rho$ vanishes, while $\phi$ converges on average to a value $R/\kappa$ and, for generic values of $\rho$, it converges to $\phi_{st}=(R-w_{2}\;\rho)/\kappa$. Let us remark that these are the noiseless or mean-field expectation values. Defining a new field $\psi=\phi-\phi_{st}$ and linearizing the dynamics as above, it is straightforward to see that this is an instance of the non-symmetric case discussed above in which the linear terms in both equations do not vanish simultaneously: DP behavior is predicted.  Observe also that, after shifting the secondary field, a standard saturation term proportional to $-\rho^2$ appears in the first equation, i.e. the competition for the available resource (water) acts as an inhibitor generating the necessary effective carrying capacity term (see main text). 

\item[{\it b)}] The case of the stochastic Levin-Segel model, as defined by Eqs.(\ref{LS_Lang1}) is a bit trickier. Observe that on average one expects (at a noiseless or mean-field level) $\phi=\rho w_2/\lambda$, which implies that when $\rho$ vanishes so does $\phi$ opening the door to non-DP scaling. However, closer inspection reveals that the noise in the equation for $\phi$ is a higher order one as it is proportional to $\sqrt{\rho\;\phi}$. It is therefore irrelevant in the renormalization group sense around the DP-like fixed point (indeed, we have verified numerically that the critical exponents of the modified version of Eqs.(\ref{LS_Lang1}) without such a term coincide, within error bars, with those of the original ones --results not shown). In this way, the equation for $\partial_{t}\phi$ becomes asymptotically a deterministic one. Then, $\phi(x)$ can be written as a function of $\rho(x,t)$, i.e.  $\phi(x,t)= F (\rho(x,t))$ where $F$ is an unspecified function which can be formally obtained by integration of the equation for $\partial_{t}\phi$. Expanding such a function in power series of $\rho$, plugging in the result in the equation for $\partial_{t}\rho$ and neglecting higher-order terms, one readily recovers Eq. (\ref{RFT}), and hence DP scaling. The field $\phi$ is just a {\it slave mode} of $\rho$ (proportional to it up to leading order), inheriting all its critical properties from it.

\item[{\it c)}] Finally, for the non-local Langevin model for Brownian bugs, it is easy to see that by performing a coarse graining in which scales below $l$ are scaled-up, one recovers effectively Eq.(\ref{RFT}). The main point is that nonlocal but finite-range interaction becomes local in the renormalization group perspective once a sufficient amount of coarse graining has been considered. See \cite{framos} for further details.
\end{itemize}


\begin{thebibliography}{99}
\bibitem{footnote} Even worse, in some cases, the described derivation of continuous approaches relying on standard field-theoretical techniques does not produce any Langevin equation, i.e. the resulting generating functional cannot be cast into any standard stochastic equation.
\bibitem{note0} In the assumption of $D$ and $D_{2}$ being constant, it is implicitly assumed that there is no intraspecific variability or other kind of heterogeneity in the distribution of these dispersal rates.
\bibitem{note3} Of course, active aggregation must be added on top of this to bring the patterns to a more realistic scenario. However, our objective is to show that there may be two complementary origins of pattern that must be integrated, one arising from stochastic demographic fluctuations and the other owing to active aggregation.
\bibitem{note4} Note that, in this case, we are using as a reference at the microscopic level the real interactions and patterns observed in Nature (see \cite{chris,chris_model_PRL,shnerbPRE} for some illustrative examples).
\end{thebibliography}
\end{document}